# Revealing Pronounced Electron-Hole Fermi Pockets in the Charge Density Wave Semimetal LaTe$_3$

T. Nakamura, Y. Fujisawa, B. R. M. Smith, N. Tomoda, T. J. Hasiweder, Y. Okada

Okinawa Institute of Science and Technology (OIST), Okinawa 904-0495, Japan.

## Abstract

Rare earth tri-tellurides (RTe$_3$) are van der Waals (vdW) coupled semimetals ideal for exploring exotic electronic phases. LaTe$_3$ is especially important for understanding the fundamental Fermiology of the RTe$_3$ family because it is non-magnetic and has a simpler charge density wave structure. In this study, we used spectroscopic-imaging scanning tunneling microscopy to measure the Landau levels of LaTe$_3$ with high energy resolution at 300 mK. These measurements were taken under varying magnetic fields up to 15 T, with fine intervals of 0.02 - 0.03 T. Our results reveal a pair of pronounced electron-hole Fermi pockets of similar sizes and evidence of electron-boson coupling in both pockets. Given the strong charge susceptibility typical of low-dimensional conductors, the interactions and instabilities driven by the electron-hole Fermi pockets could be a basis for searching unexplored quantum phases in other antiferromagnetic RTe$_3$ compounds.

**Introduction**

Low-dimensional semimetals offer a promising platform for exploring the emergence of exotic electronic phenomena. These materials host a variety of quantum states, including topologically nontrivial phases, excitonic insulators, unconventional superconductivity, and charge density wave (CDW) states that intricately interact with lattice, spin, and orbital degrees of freedom [1,2,3,4,5,6]. Among these, the rare earth tri-telluride ($RTe_3$) family has gathered significant attention, with recent studies highlighting its rich electronic and magnetic properties [7,8,9,10,11,12,13,14,15]. From a fermiology perspective, the $5p_x$ and $5p_z$ orbitals of the Te square net are the relatively simple starting point. However, necessary band folding arises due to the presence of the block layer (**Fig. 1a-c**). Additionally, the band folding due to multiple CDW formations and associated momentum-dependent gap openings results in the complexed Fermi surface and associated rich electronic instabilities in $RTe_3$ (**Fig. 1c**) [14,16,17,18,19,20,21]. Therefore, isolating the key Fermi surface features is a crucial step in establishing a guiding principle for searching unexplored novel low-temperature electronic states in $RTe_3$ compounds.

$LaTe_3$ occupies an important position within the $RTe_3$ family, as highlighted by the orange-shaded region in the schematic phase diagram (**Fig. 1d**). The absence of 4f electrons renders $LaTe_3$ non-magnetic, and only a single CDW (indicated by CDW1 in **Fig. 1d**) is known to occur around 600 K. This makes $LaTe_3$ an ideal system to investigate the nature of parent Fermi surfaces realized by the CDW1 phase, which hosts rich phases emerge at low temperature in $RTe_3$, including CDW2, exotic magnetism, and possible hidden intertwined phases (**Fig. 1d**). Previous studies have proposed the Kramers nodal line near the Brillouin zone (BZ) boundary and large magnetoresistance has also been reported in $LaTe_3$ [22,23]. These findings suggest a semi-metallic band structure with effective charge compensation between electrons and holes. Additionally, a recent optical study revealed an axial Higgs-like bosonic mode with a characteristic energy scale of ~10 meV, which is suggested to couple with electrons near the BZ boundary [24]. Despite previous studies, a comprehensive understanding of many-body effects, density wave formation, and electron-hole nesting conditions remains incomplete. Landau level (LL) spectroscopy, performed using a scanning tunneling microscope (STM), offers a powerful method to probe these phenomena [25].

In this report, we present the first LL spectroscopy of $LaTe_3$, using spectroscopic STM. The temperature was fixed at 300 mK under controlling magnetic fields (*B*) up to 15 T, with 0.02~0.03 T interval. The observed LL dispersion reveals the coexistence of electron and hole bands with comparable Fermi pocket sizes. Additionally, we observe a pronounced sharpening of the LLs and the emergence of a dispersion kink near the Fermi level ($E_F$). These findings in $LaTe_3$ provide new insights into the essential Fermiology of the RTe$_3$ family.

## Results and Discussions

### Methods

High-quality single crystals of LaTe$_3$ were grown from a binary melt, as described in previous studies [26,27,28]. Scanning tunneling microscopy (STM) was performed on the freshly cleaved surfaces of LaTe$_3$ at 300 mK. All measurements employed an electrochemically etched tungsten (W) tip, conditioned on an Au(111) surface. Standard lock-in techniques were used to acquire differential conductance (d$I$/d$V$) spectra, with a modulation frequency of 961 Hz. All data presented in this report were obtained without any observable change in tip conditions.

### Observation of CDW1 in LaTe$_3$

Before delving into the LL spectroscopy results, we confirm that the magnetic field does not alter the CDW modulation. Cleaving the sample exposes the Te square net. The typical surface topography at $B$ = 0 T, captured over a 50 nm × 50 nm area, is shown in **Fig. 1e**. Magnified images reveal the atomic resolution of the topmost Te layer, with the one-dimensional CDW running along the c-axis (**Fig. 1f**). Note that the crystallographic convention designates the a- and c-axes as the in-plane directions. To extract the periodic structure, a 2D Fourier transform (FT) of the 50 nm × 50 nm area was performed (**Fig. 1g**). The Bragg peaks, $q_a$ and $q_c$, corresponding to the LaTe$_3$ unit cell are observed (the black circles corresponding the black square in **Fig. 1b**). The CDW wave vector $q_{CDW} \approx (5/7)q_c$ is marked by an orange circle. Several additional peaks, indicated by arrows in **Fig. 1h**, correspond to CDW propagation vectors formed by linear combinations of $q_c$ and $q_{CDW}$ [29,30,31]. For instance, the CDW propagation vector along the horizontal direction in **Fig. 1c**, labeled as $q'_{CDW}$, follows the relation $q'_{CDW}$ = $q_c$ - $q_{CDW}$ (see arrows in **Fig. 1h**). The excellent overlap of the FT line profiles along the $q_c$ direction at $B$ = 0 T and $B$ = 15 T confirms that the CDW in LaTe$_3$ is independent of the applied magnetic field. The magnetic field-independent CDW property seen in LaTe$_3$ is coherently interpreted as CDW1 on the established phase diagram (**Fig. 1d**).

### Appearance of Landau level (LL) in magnetic field

While magnetic field-induced deformation of the CDW is absent, the emergence of LLs becomes apparent at $B$ = 15 T. **Fig. 2a** and **2b** show the spatial evolution of the differential conductance (d$I$/d$V$) along the direction indicated by the black arrow in **Fig. 1e**, measured at $B$ = 0 and $B$ = 15 T, respectively. To suppress any remaining weak signals from the standing wave patterns, spatially averaged d$I$/d$V$ spectra were obtained (**Fig. 2c**). At $B = 0$, the spectrum exhibits a dip structure within an energy range of ± 20 meV, centered around the Fermi level ($E_F$), as indicated by the black arrows. This energy scale is significantly smaller than the characteristic CDW1 soft-gap (~ 400 meV). Since no CDW1 deformation occurs upon applying a magnetic field (**Fig. 1h**), additional band folding is unlikely to account for the fine peak structures observed near $E_F$ at $B = 15$ T. Therefore, the additional peaks in the d*I*/d*V* spectra at $B = 15$ T are indicative of LL formation. The relatively homogeneous LL-derived peaks (seen along the black arrow in **Fig. 2b**) suggest minimal potential variation across the sample, ensuring a reliable platform for LL spectroscopy [32,33].

**Coexistence of electron and hole bands near $E_F$**

For LL analysis, the initial task is to clarify the energy positions of the LLs. Taking the second derivative of the differential conductance spectra ($- d^3I/dV^3$) is a well-established method for enhancing the visibility of peak and hump positions corresponding to LLs [34,35,36,37,38]. To demonstrate this in our case, we compare the raw d*I*/d*V* (black curve) with the $- d^3I/dV^3$ (red curve) at $B = 15$ T (**Fig. 2d**). As shown, subtle LL signals are effectively amplified in the $- d^3I/dV^3$ spectrum. The vertical dotted lines in **Fig. 2d** mark the extracted LL energy positions, whose magnetic field-dependent shift is crucial to track. The magnetic field dependence of the LLs is investigated by focusing on spatially averaged d*I*/d*V* spectra in the field range between 12.6 T and 15.0 T, with intervals of 0.02 - 0.03 T. The corresponding $- d^3I/dV^3$ image plot is presented in **Fig. 2e**. As indicated by the dashed lines in **Fig. 2e**, the LL dispersion with a positive slope (from upper right to lower left) is observed, particularly for the occupied states. Notably, rather sharp LLs are evident within a ~ 20 meV energy window from $E_F$. Within this energy window, additional LL dispersions with a negative slope also become evident (from lower right to upper left in **Fig. 2e**). These observations suggest the coexistence of sharp LLs from electron and hole bands near the $E_F$ (as indicated by the arrow on the top axis in **Fig. 2d**). See **Appendix A** for the details to extract the energy positions of the LLs.

**Dispersion anomaly near $E_F$**

For quantitative analysis, the next essential task is assigning independent indices to the LLs of the two bands. By tracking the continuous energy shifts of LLs as the magnetic field changes, the extracted LL dispersions near $E_F$ are shown in **Fig. 3a**. To ensure accurate tracking of the LLs, the dispersions are overlaid on the $- d^3I/dV^3$ image plot (**Fig. 3b**). See **Appendix A-2** and **A-3** for the details of peak search procedure. When LLs are indexed correctly, the dispersion remains largely unchanged with varying magnetic fields, aside from small Zeeman energy effects. This situation realizes that all LLs under various magnetic fields can be collapsed onto a single field-independent dispersion [37,39]. By imposing this constraint in the nonmagnetic LaTe$_3$ case, the LLs' indices for the electron and hole bands are uniquely determined. Using the lowest-energy LL for the electron band ($n_{e0}$) and the highest-energy LL for the hole band ($n_{h0}$), $n_{e0}$ = 39 and $n_{h0}$ = 111 are estimated (**Fig. 3c-d**). The obtained physical parameters, through the indexing process, are summarized in **Table 1.** No inconsistency with previous studies is seen [22,40], pointing out that LL indexing in our study is successful. Further details of the analysis are provided in **Appendix B**. For clarity, the electron and hole bands are referred to as the β and γ bands, respectively.

**Self-energy (Σ) based analysis and coupling constant (λ)**

A dispersion kink near $E_F$ in the β-band is evident, as highlighted by the dashed circle in **Fig. 3c**, appearing symmetrically in both occupied and unoccupied states. This kink likely originates from a many-body interaction-induced deviation from the bare band dispersion, corresponding to the real part of the self-energy, Re(Σ) (**Fig. 4a**). Here, Δ represents the experimental deviation from the bare band dispersion, serving as a basis for theoretical Re(Σ) (**Fig. 4a**). **Figure 4b** illustrates the energy dependence of Δ. In this figure, Δ values obtained from a peak search on $-d^3I/dV^3$ without fitting (red symbols) overlay with peaks from multiple Lorentzian-based fittings on $dI/dV$ (gray symbols). The close agreement between these data sets, as shown in **Fig. 4b**, supports the reliability of using multiple Lorentzian-based fittings to extract Landau level (LL) line widths, further discussed below.

To investigate any asymmetry in the kink between unoccupied and occupied states, we plot experimental |Δ| values and its linear interpolation (Lerp |Δ|) as a function of |$E - E_F$| (**Fig. 4c**). The observed symmetry supports the interpretation of experimental |Δ| as a valid measure of Re(Σ). For simplicity, **Fig. 4c** selectively presents Δ values extracted directly from a peak search on $-d^3I/dV^3$ without fitting. The range of |Re(Σ)| up to ~ 20 meV around $E_F$ aligns with the energy range associated with the major phonon density of states in this system [24]. Focusing on the β-band, we estimate the electron-phonon coupling constant λ = 0.19 ± 0.01, based on the linear slope of Δ fitted around $E_F$ (**Fig. 4c**), indicating weak coupling. Further details on the calculation of λ are found in **Appendix C**.

**Landau level energy widths**

We now discuss potential many-body effects on the γ-band. While the many-body interaction in the β-band is well-documented, its presence in the γ-band is challenging to establish. This difficulty arises from the absence of clearly observable LLs beyond the kink energy (± 10 meV from $E_F$) in the γ-band (**Fig. 3a, b**). Theoretically, the quasiparticle lifetime is related to the imaginary part of the self-energy, Im($\Sigma$), which is connected to Re($\Sigma$) through the Kramers-Kronig (KK) transformation. Notably, in **Fig. 4d**, the KK-transformed Lerp $\Delta$ for the β-band aligns closely with the experimentally observed LL half-width at half maximum (HWHM), obtained via multiple Lorentzian-based fittings on d$I$/d$V$ [see Eq. (1) and **Appendix A-3**]. Thus, empirically, the LL HWHM serves as a measure of the quasiparticle lifetime $|\text{Im}(\Sigma)|$. Significant LL broadening across a characteristic energy range could lead to undetectable LLs, suggesting electron-boson coupling in the γ-band where LLs remain unresolved for $|E - E_F|$ > 10 meV. Interestingly, while LL line widths are comparable between the β- and γ-bands within $|E - E_F|$ < 10 meV (**Fig. 4d**), the γ-band shows more substantial fading outside this range, possibly indicating a higher coupling constant than in the β-band. However, the intrinsic difficulty in defining the bare band for the γ-band makes this speculative. For further details on self-energy analysis in LL spectroscopy, see **Appendix C**.

**Electron-hole pockets near the zone boundary**

In non-magnetic LaTe$_3$, it is natural to assume that $\Sigma$ is field-independent. This is corroborated by the dip structure consistently observed within an energy range of ± 20 meV from $E_F$ at $B$ = 0 T (**Fig. 2c**). Therefore, information from LL spectroscopy should be used to understand $B$ = 0 band structure. While the LL dispersion provides Fermi pocket size, its position in $k$-space is intrinsically unresolved sorely from LL spectroscopy. To search the relevant Fermi pocket positions, the underlying complex Fermi surfaces are first mapped in $k$-space, based on a tight-binding model [15,21] (**Fig. 4e**). For simplicity, the ungapped Fermi surface portion near BZ boundary is selectively shown in **Fig. 4e**. The black curve includes Te 5p derived two bands in Te-square net and their folding due to periodicity in LaTe blocking layer (same as **Fig. 1c**). These bands are folded by CDW1 formation (see **Fig. 1**), and consequent shadow bands are additionally shown by green lines in **Fig. 4e**. Notably, the Fermi pocket sizes drawn from LLs (blue and red curves in **Fig. 4e-g**) agree well with electron and hole bands centered at X point. It should be stressed that our finding successfully isolates the essential portion of Fermi pockets out of complexity [41,42,43,44]. Intriguingly, a non-equilibrium CDW phase with a propagation vector, $q_{pulse}$ ~ 0.29, is discussed in LaTe$_3$ induced by a laser pulse [20]. The absolute value of this hidden CDW propagation vector approximately corresponds to the Fermi surface nesting vectors, which are required to overlap electron-hole Fermi surface pockets efficiently (see the arrow in **Fig. 4f**). While this none-equilibrium CDW phase is hidden in the equilibrium state, the relevant underlying Fermi surface instability can be roughly speculated based on the electron-hole Fermi pockets presented in this study.

**Significance**

Since the Fermi surface characteristics are expected to be similar across all $RTe_3$ compounds, the insights gained from $LaTe_3$ are likely relevant to deepen understanding of rich electronic emergence in the entire $RTe_3$ family [45]. A particularly interesting case is considering $CeTe_3$, as La and Ce are adjacent in the periodic table and possess similar ionic radii (**Fig. 1d**). Furthermore, amongst other R cases, the $4f^1$ states in Ce ion host stronger coupling with conducting electrons on Te square net [4, 14,21,26,27,28,46,47,48]. Out of the pronounced electron-hole Fermi pockets, searching magnetic CDW characterized by multiple-$q$ states is an interesting perspective via external perturbations, which is crucial in searching for topological magnetism [49].

**Summary**

In summary, we have presented Landau level (LL) spectroscopy of $LaTe_3$ at 300 mK using scanning tunneling microscopy. Among the coexisting intricate Fermi surfaces, the successful isolation of essential electron-hole Fermi pocket portions is a major finding of this study. The relevant interactions and instabilities driven by electron-hole Fermi pockets in $LaTe_3$ could be a basis for searching the unaccused quantum state of matter in other antiferromagnetic $RTe_3$ compounds.

**Acknowledgment**

## Appendix A: Analyzing Landau Level (LL)

In this appendix, we provide a detailed, step-by-step procedure for extracting the peak position and width of LLs employed in this study. These steps include the process of getting relatively smooth -$d^3I/dV^3$ along the energy axis (**A-1**) and the extraction of LL energies and width in characteristic two energy regions (**A-2** and **A-3**). Using multiple approaches to extract the peak position and width of LLs, the validity of the conclusions in this paper is checked in this study.

### A-1. The process to get -$d^3I/dV^3$

The second derivative of the -$dI/dV$ spectra (-$d^3I/dV^3$) is used to determine the energy of LLs via a Savitzky-Golay (SG) filter. This is a type of moving-average filter [51], which requires uniformly spaced experimental data. In this study, the energy sampling interval for the $dI/dV$ spectrum is typically 0.2 to 0.3 meV. To calculate the -$d^3I/dV^3$ at the *m*-th data point, ± 21 data points around the *m*-th point of $dI/dV$ are approximated by a fourth-degree polynomial $f$ described by $f_m(x) = \sum_{i=0}^{4} a_{mi} x^i$, where $a_{mi}$ are fitting parameters. Here, *x* = 0 corresponds to the central *x* position in the local fitting function $f_m(x)$. Then, SG-filtered value of $dI/dV$ at the *m*-th point is obtained by $f_m(0) = a_{m0}$. Following this process, the *m*-th data point of -$d^3I/dV^3$ is described as $-d^2/dx^2 f_m(0) = -2! a_{m2}/(\Delta x)^2$ where $\Delta x$ is interval of data points. By comparing $dI/dV$ and -$d^3I/dV^3$ carefully (**Fig. 2d**), the number of data points used for local fitting was carefully chosen not to create the artifacts in our analysis.

### A-2. High throughput determination of LL energies

To obtain LL energies, without involving the curve fitting process, we can identify local maxima of -$d^3I/dV^3$ by tracking differences between adjacent data points. Having a relatively smooth -$d^3I/dV^3$ curve (as in **Appendix A-1**) is a crucial step in this high throughput peak search process. Despite the simplicity of this method, it effectively tracks peaks. In **Fig. 3b**, successful overall peak tracking is demonstrated by overlaying the extracted peaks on the -$d^3I/dV^3(E,B)$ image plot. The extracted LLs display continuous shifts with changing magnetic fields (**Fig. 3a**), which are physically consistent. Here, at the crossing points of each LL dispersion (**Fig. 3a**), LLs from both electron and hole bands are assumed to coexist.

### A-3: Lorentzian-based fitting for LLs in two energy regions

While the LL energy is obtained directly from -$d^3I/dV^3$ without curve fitting, estimating the LL line width requires a fitting process on the $dI/dV$ data. To accurately evaluate the half width at half maximum (HWHM) of the LLs, we first subtract the background by removing the $dI/dV$ spectrum at *B* = 0 from the d$I$/d$V$ spectrum measured under a finite magnetic field. After this process, the background subtracted d$I$/d$V$ were fitted using the following function:

$$I(E) = b_0 + b_1 E + b_2 E^2 + \sum_n L_n(E) \qquad (1)$$

This function consists of an empirical quadratic background with multiple Lorentzian functions. $b_0$, $b_1$, and $b_2$ are the parameters of the quadratic background function on tunneling spectra. The $L_n(E)$ is the single Lorentzian function for $n$-th LL given by $L_n(E) = h_n \frac{w_n^2}{(E-E_n)^2 + w_n^2}$, where $h_n$ is the peak height, $w_n$ is the HWHM, and $E_n$ is the energy level for $n$-th LL.

In this study, we considered two energy regions with distinct approaches to extract HWHM. One region is away from $E_F$ ($|E - E_F|$ > ~20 meV), where each LLs from the β-band appears as separated peaks at a fixed magnetic field (see outside of the shaded area in **Fig. 2e**). In this energy region, without further process, d$I$/d$V$ curves at fixed magnetic field are fitted by Eq. (1). On the other hand, we noticed that a similar fitting approach is not applicable for energy regions near $E_F$ (see inside of shaded area in **Fig. 5a**). In this region, several LLs are difficult to detect as separated peak positions. To specify this issue further, the image plot of the d$I$/d$V$($B$,$E$)- d$I$/d$V$(0,$E$) for $|E - E_F|$ < 20 meV (**Fig. 5b**) is compared with the background subtracted d$I$/d$V$ curve at a fixed magnetic field of 14.96 T (**Fig. 5c**), which corresponds to the orange horizontal line in **Fig. 5b**. Relying on the continuum of LLs seen in **Fig. 5b**, the expected existence of LLs' energies are indicated by dotted vertical lines in **Fig. 5c** (blue for electrons and red for holes). In **Fig. 5c**, significant overlap of multiple LLs results in unresolved separation of peaks from multiple LLs. Thus, evaluating the width of each LL for $|E - E_F|$ < 20 meV is not a straightforward technical process, motivating the development of an alternative approach.

We focused on the evolution of d$I$/d$V$ spectra along diagonal lines for $|E - E_F|$ < 20 meV (**Fig. 5b**). It should be stressed that the diagonal line cut is positioned to follow the valley between adjacent LLs from the electron band (see the blue line in **Fig. 5b**). The purpose of this process is to minimize the contribution of the electron band to create fine energy dependence of d$I$/d$V$, which is beneficial to analyze LLs from hole band as independent as possible. The obtained diagonal line cut profiles are then projected onto the energy axis, allowing for the extraction of LL line widths [$w_n$ in Eq. (1)] by fitting with multiple Lorentzian functions (**Figs. 5d**, **e**). The effectiveness of this method is demonstrated by the comparison between **Fig. 5c, d,** and **e**. Note that the LL line width extracted via the diagonal line cut requires the following correction to properly project the LL line width onto the energy axis at the fixed magnetic field. In the case of LLs from the electron band, the correction to the width of LLs from electron (hole) band $w_{e(h)}$ is

$$w_{corr} = w_{e(h)}\left(1 + \frac{\tan\theta_{h(e)}}{\tan\theta_{e(h)}}\right) \quad (2).$$

Here, $\tan\theta_{h(e)}$ represents the slope of LLs from the hole (electron) band (see **Fig. 5b**). By focusing on different valleys, we similarly estimated HMHW for all relevant LLs from 14.6 T to 15.0 T for $|E - E_F| < 20$ meV. Similarly, a diagonal cut is used to analyze the LL line width from the electron band (see the red line in **Fig. 5b**). In **Fig. 4d**, all LL HMHW from the electron band for $|E - E_F| < 20$ meV are plotted $w_{corr}$ based on Eq. (2).

The $E_n$ values obtained via the methods outlined in **Appendices A-2** and **A-3** are overlaid in **Fig. 4b**. The reasonable overlap confirms that the two methodologies, across two distinct energy regions, consistently capture the underlying continuous trend. The diagonal line-cut method profiles across different magnetic fields, where the field dependence of the LL wavefunction's spatial extension could influence the LL widths [33]. However, given the small and limited range of magnetic fields considered, we disregard the effect of the changing magnetic field on the HMHW in our analysis. Within this narrow $B$ - window, the energy dependence of the HWHM reasonably aligns with the quasiparticle lifetime dependence [Im($\Sigma$)], further discussed in **Appendix C**.

**Appendix B: indexing LLs**

Indexing LLs becomes an important task in this study. Since the band edges are far from $E_F$ in our case, the detectable near $E_F$ LLs should be assigned by high indices without prior knowledge of the energy position of the lowest indexed LL. This results in the requirement of a nontrivial task of indexing LLs, which includes modeling of LL dispersion (**B-1**) and LL index determination (**B-2**). The calculation of physical properties from LL spectroscopy is also explained in (**B-3**).

**B-1. LL dispersion**

From the quantized energy levels $E_n$, the corresponding wave vector $k_n$ can be extracted using the formula, $k_n = \sqrt{\frac{2eB}{\hbar}\left(n + \frac{1}{2}\right)}$, where the LL index $n$ is assigned. Here, $\hbar$ and $e$ are reduced Plank constant and elementary charge, respectively. The formula is derived from the Lifshitz-Onsager quantization condition, which is a universal approach in the scope of semiclassical theory. When indices $n$ are correctly determined, STM can extract a discrete version of the energy-momentum dispersion relation ($E_n$ vs. $k_n$) at a fixed magnetic field. Here, the deformation of underlying band dispersion induced by $B$ is negligible in non-magnetic material $LaTe_3$, and the Zeeman effect can be conventionally small. Therefore, the set of LL energy and corresponding momentum ($E_n$ vs. $k_n$) under various $B$ should be collapsed onto a single underlying band. Notably, this collapsing can be employed as a constraint to determine LL indices [34,36,39]. For determining the LL indexes, $n_{e0}$ ($n_{h0}$) is assigned as the fitting parameters representing the lowest (highest) energy level among the observable LLs used for fitting. Thus, the main challenge to determine LL indices is figuring out $n_{e0}$ and $n_{h0}$.

**B-2. Bare band and indexing LL**

To quantify the collapse of LLs onto the underlying band, we phenomenologically introduced a following dispersion as a function of the wavenumber $k$,

$$E(k) = c_0 + c_2 k^2 + c_4 k^4 \quad (3).$$

Here, $c_0$, $c_2$, and $c_4$ are fitting parameters. When LL dispersion near the band top or bottom is considered, introducing a simple parabolic band would be sufficient with $c_4 = 0$ (with either $c_2 > 0$ or $c_2 < 0$). On the other hand, when LL dispersion away from band edges is considered, introducing the quartic component (in addition to the parabolic component) would be useful to represent the underlying coexistence of electron- and hole-like band curvatures. In our case, we recognize that the modeling the bare electron band (β-band) requires opposite signs for $c_2$ and $c_4$ (i.e., $c_2 > 0$ and $c_4 <$ 0). Based on the band curvature point of view, this means the requirement of a minor hole-like contribution in addition to a dominant electron-like nature. Since LLs for γ-band is within the kink energy window (**Fig. 3d**), the physical reasoning for modeling dispersion is unclear. However, this unambiguity does not have a major impact on the main scope of this study. In this study, the parabolic function (with $c_4 = 0$) is simply employed empirically to capture field-independent LL dispersion near $E_F$. By adjusting the offset indices $n_{e0}$ and $n_{h0}$ (see **Fig. 2d** and **Fig. 3a-b**) the residual sum of squares between the experimental $E_n$ values and the underlying band energy $E$ [Eq. (3)] is calculated using the following formula.

$$s = \sum_{n=n_{e(h)0}}^{n_{e(h)0}+\text{num. of LLs}} \sum_{k_n} \{E(k_n) - E_n(k_n)\}^2 \quad (4)$$

Based on Eq. (4), the minimum $s$ is explored and plotted in **Fig. 6d** or **6e** as a function of $n_{e0}$ (or $n_{h0}$), where $c_0$, $c_2$, and $c_4$ are self-consistently optimized for each $n_{e0}$ (or $n_{h0}$). To compute $s$, data within ± 30 meV of $E_F$ are excluded to avoid the intrinsic challenges in formulating the energy-dependent self-energy for the electron band (see **Appendix B**). The optimal offset indices were identified as $n_{e0} = 39$ for the electron band and $n_{h0} = 111$ for the hole band (see **Figs. 6d, 6e**, and **Figs. 3c, 3d**). For instance, a smaller $s$ is evident for $n_{e0} = 39$ compared to the other two cases (**Figs. 6a-c**).

### B-3. Band parameters obtained

Fermi velocity ($v_F$) and effective mass ($m^*$) were calculated from the fitting curves using the following equations:

$$v_F = \frac{1}{\hbar} \frac{\partial E(k)}{\partial k}\bigg|_{k=k_F} \quad (5)$$

$$\frac{1}{m^*(k)} = \frac{1}{\hbar^2} \frac{\partial^2 E(k)}{\partial k^2} \quad (6).$$

From the LL dispersion, the Fermi wave number ($k_F$), Fermi velocity ($v_F$), and the area of the Fermi surface ($A_{FS}$) relative to the area of the Brillouin zone ($A_{BZ}$) were also calculated using $k_F$ and $A_{FS}/A_{BZ}=\pi k_F^2/A_{BZ}$, respectively. Here, $A_{BZ} = 2.027$ Å$^{-2}$ is calculated from the lattice constants of LaTe$_3$ ($a = 4.4045$ Å and $c = 4.421$ Å) [40]. The obtained values are consistent with those from transport and ARPES literature (see **Table 1**).

# Appendix C: Self-energy ($\Sigma$) based analysis

This study analyzes the LL dispersion based on the self-energy $\Sigma$ [52]. (**C-1**) describes the methodology for extracting the coupling constant $\lambda$ based on Re($\Sigma$). In (**C-2**), the Kramers-Kronig (KK) transformation is explained to interpret the LL line widths based on Im($\Sigma$).

## C-1: Coupling constant $\lambda$

The following equation defines the electron-phonon coupling constant.

$$\lambda = \left.\frac{\partial \text{Re}(\Sigma)}{\partial E}\right|_{E=0} \quad (7)$$

$\lambda$ and $\Sigma(k, E)$ are coupling constant and self-energy, respectively. As shown in **Fig. 4b**, Re($\Sigma$) is approximately linear near $E_F$. Instead of differentiating Re($\Sigma$), near $E_F$ is fitted with linear function to get $\lambda$ (**Fig. 4c**). The error for $\lambda$ is from the fitting error based on Eq. (7),

## C-2: Kramers-Kronig (KK) transformation

By defining $\Sigma(k,\omega)$ as the self-energy and $P$ as the Cauchy principal value, Im($\Sigma$) can be obtained from the Kramers-Kronig (KK) transformation of Re($\Sigma$). When the self-energy $\Sigma$ is a continuous and differentiable complex function, the real and imaginary parts of the self-energy are related through the following KK relation.

$$\text{Im}\Sigma(k,\omega) = -\frac{1}{\pi} P \int_{-\infty}^{+\infty} \frac{\text{Re}\Sigma(k,\omega')}{\omega-\omega'} d\omega' \quad (8)$$

It should be stressed that the correspondence between the experimental $\Delta$ (and Lerp $\Delta$) and Re($\Sigma$) is demonstrated (**Fig. 4a**). Also, $\Delta$ nearly saturates to zero outside the energy window |$E$ - $E_F$| ~ 30 meV (see **Fig. 4c**). Therefore, replacing Re($\Sigma$) to Lerp $\Delta$, with integrating energy range restricted between − 80 meV and + 45 meV can be assumed to provide Im($\Sigma$), following

$$\text{Im}\Sigma(k,\omega) \sim -\frac{1}{\pi} P \int_{-80\,meV}^{+45\text{meV}} \frac{\text{Lerp }\Delta(\omega')}{\omega-\omega'} d\omega' \quad (9).$$

The potential relationship between LL linewidths and quasiparticle lifetimes, which should be reflected by Im($\Sigma$), is not a physically trivial matter. However, the KK transformation of Lerp $\Delta$ aligns well with the experimental LL HWHM (see lines and symbols in **Fig. 4d**). Therefore, within the empirical framework of our study, the energy dependence of LL linewidths can be interpreted as representing Im($\Sigma$) deduced from Eq. (9).

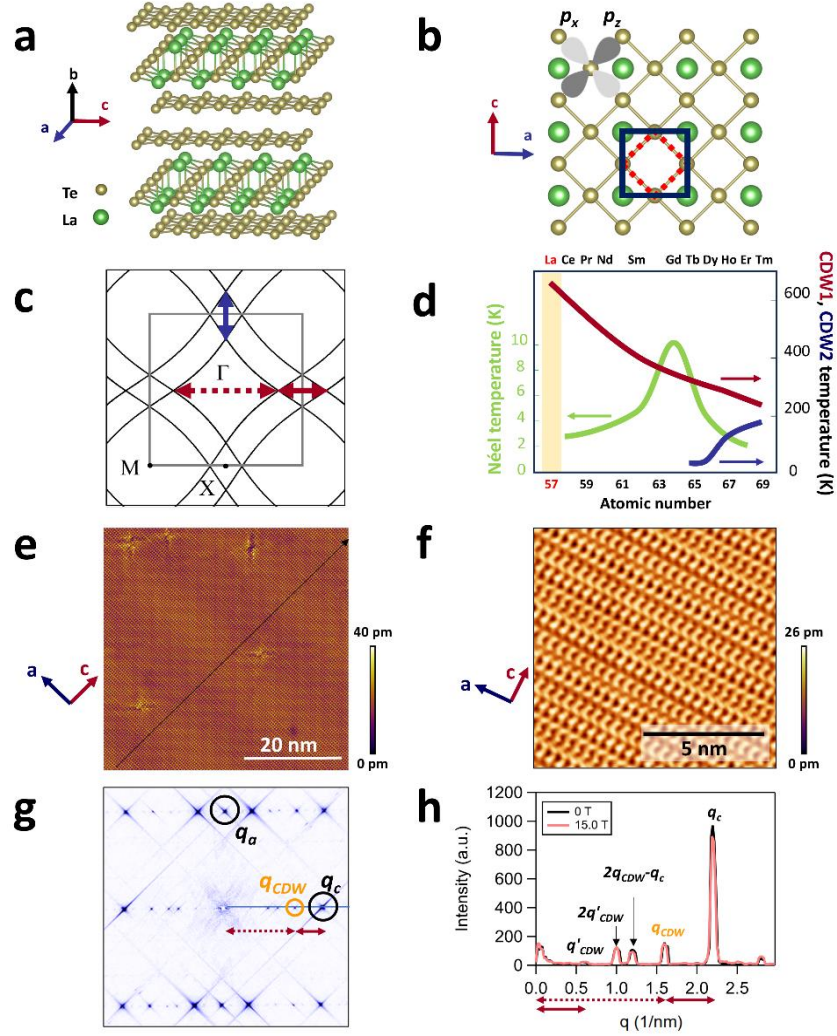

**Figure 1**

(a) Side view of the LaTe$_3$ crystal structure. LaTe slabs are sandwiched by Te square net sheets, which are weakly coupled by van der Waals (vdW) forces. (b) Top view after exfoliation. The topmost (Te) and second (La) layers are highlighted. The lateral unit cell expands (black solid square) compared to the free-standing Te net (red dashed square) due to the underlying La sheets. (c) Simplified Fermi surface schematic, with the diagonal 5p$_x$ and 5p$_z$ bands with additional folded bands by the Brillouin zone [black solid square in (b)]. This schematic ignores CDW-driven band folding and associated gap opening. (d) Phase diagram of RTe$_3$, showing CDW phases with their higher (CDW1, red curve) and lower (CDW2, blue curve) transition temperatures (right axis). With a green curve, the antiferromagnetic transition temperatures are also indicated (left axis). (e) STM topography of LaTe$_3$ (50 nm × 50 nm, -90 mV / 2.0 nA) shows the unidirectional CDW1 state. (f) The zoomed-in STM image with atomic resolution (10 nm × 10 nm, -15 mV / 200 pA). (g) FT pattern of (e), with Bragg peaks (black) from the unit cell and CDW1 (orange). (h) Line profiles of the FT pattern in (g) along q$_c$ at 0 T (black) and 15.0 T (pink), showing Bragg (nearest neighbor Te-Te bonding) and CDW1 peaks at 2.2 nm$^{-1}$ and 1.6 nm$^{-1}$, respectively, with additional peaks from wavevector mixing.

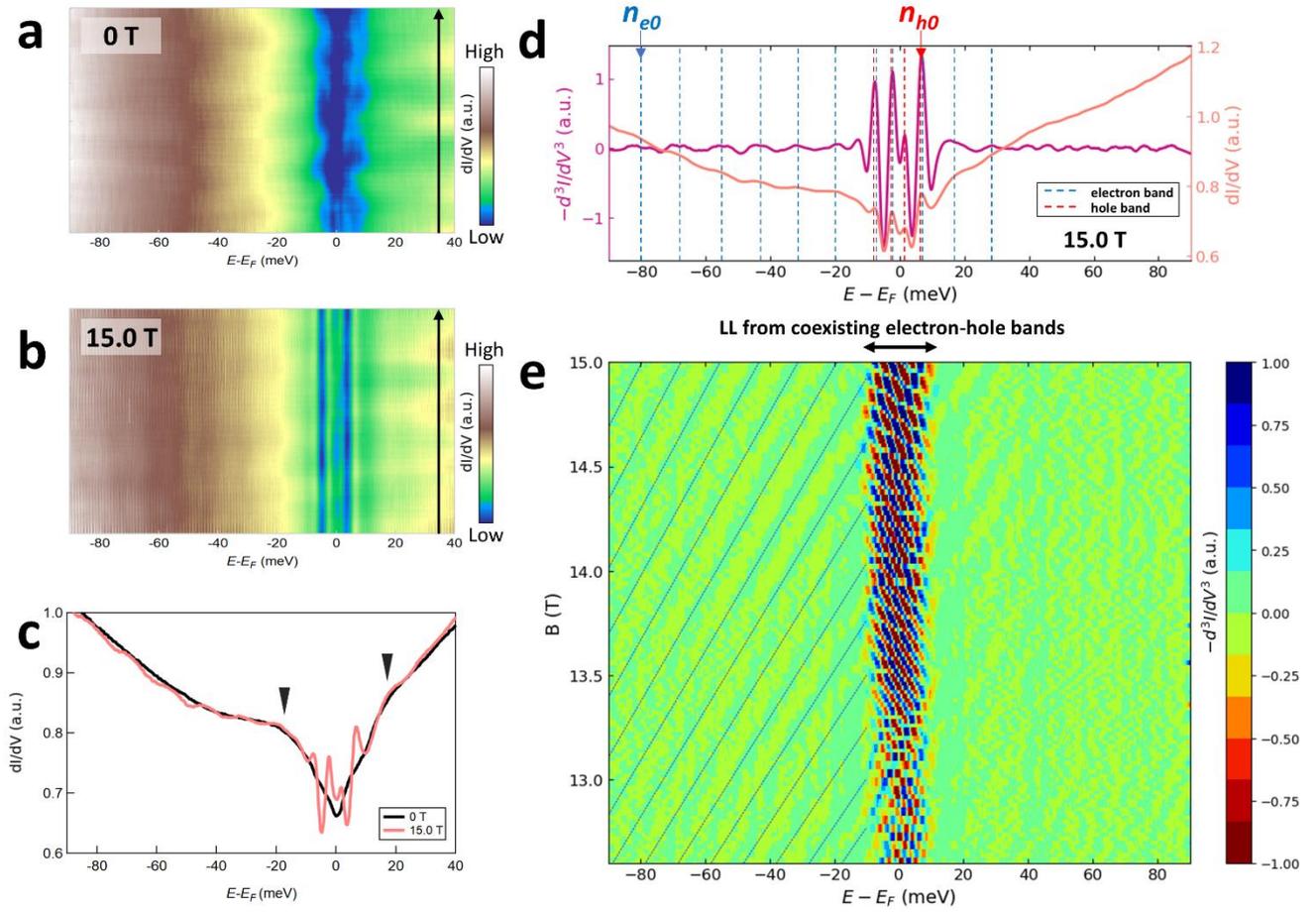

**Figure 2**

(a)-(b) Image plot of d*I*/d*V* spectra along the line with the arrow indicated in **Fig. 1(e)** at (a) 0 T and (b) 15 T. (c) Spatially averaged d*I*/d*V* spectra from **Fig. 2(a)** (black) and **Fig. 2(b)** (pink). (d) Comparison of spatially averaged d*I*/d*V* (pink) and -d$^3$*I*/d*V*$^3$ (purple). $n_{e0}$ and $n_{h0}$ represent the offset LL indexes for electron and hole bands (see main body for details). (e) Image plot of -d$^3$*I*/d*V*$^3$ as a function of energy and magnetic field. Dashed lines highlight the LLs in the occupied states from the electron band (see main text).

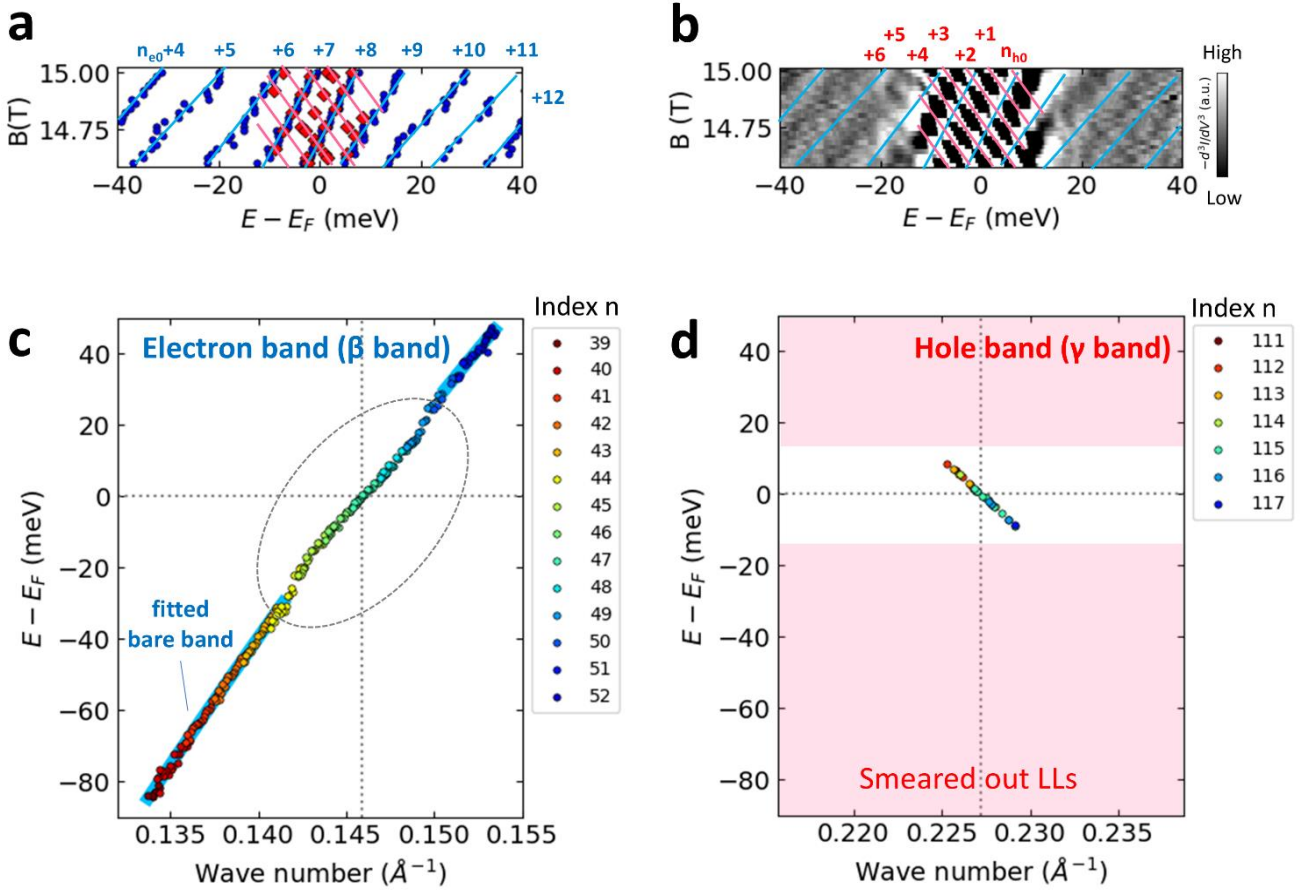

**Figure 3**

(a) LLs near $E_F$. Blue (red) markers and guidelines show LLs from the electron (hole) band. (b) Electron and hole band guidelines overlaid on the magnified image from **Fig. 2(e)**. Indexes are indicated on the top axis, using offset values $n_{e0}$ and $n_{h0}$ in (a) and (b). (c)-(d) Energy-momentum dispersions from LLs for (c) electron and (d) hole bands, with optimized $n_{e0}$ = 39 and $n_{h0}$ = 111 (see **Appendix B-3** for details). In (c), the solid blue curve represents the bare band obtained by fitting, excluding points in the kink portion (see dashed ellipse). The bare band dispersion is from Eq. (3) (see **Appendix B-2** for further details). The LL energy positions in (a)-(d) are from a local maximum of smoothly filtered $-d^3I/dV^3$ without introducing a curve fitting process (see **Appendix A-2** for details).

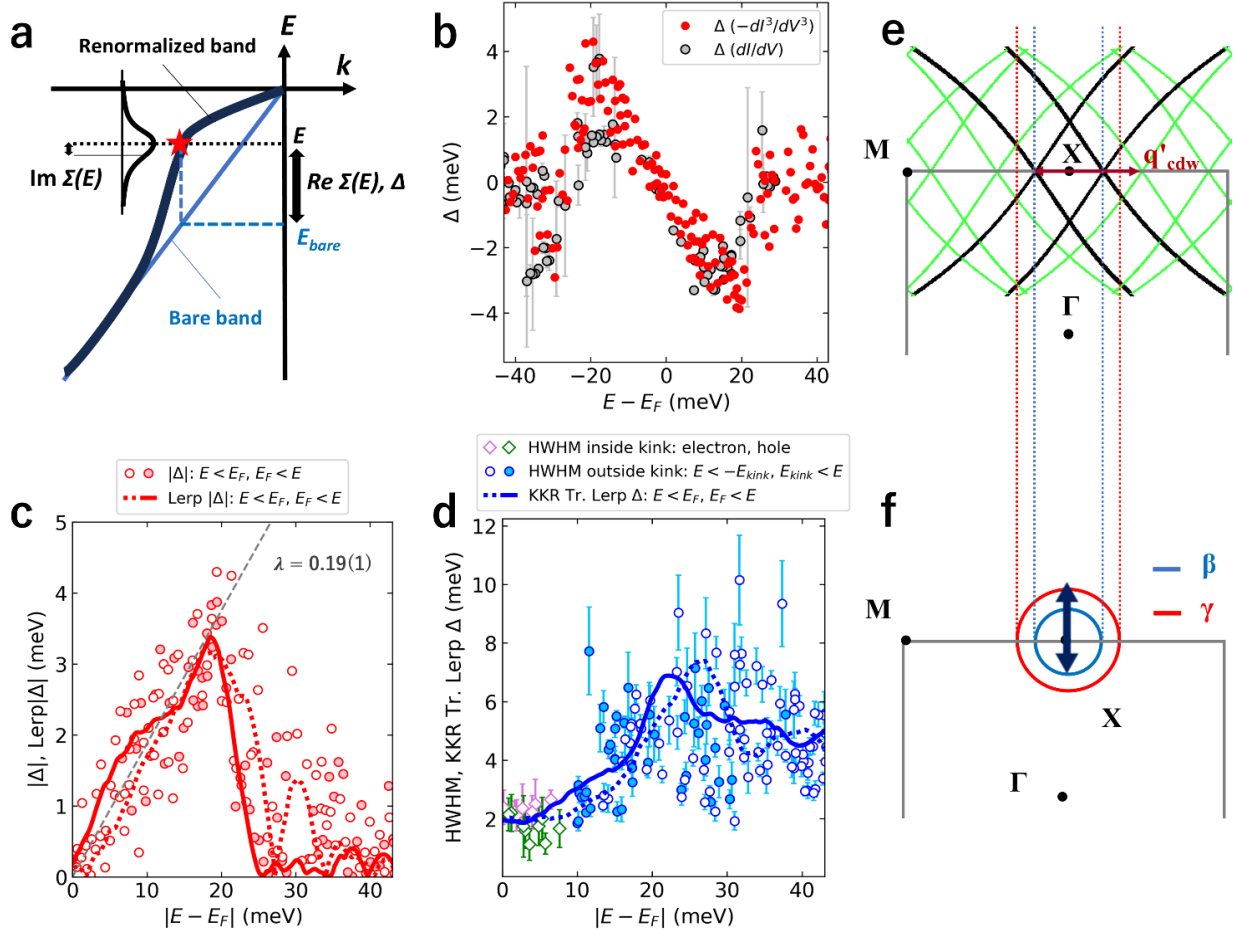

**Figure 4**

(a) Schematic of the renormalized (thick black) vs. bare (thin blue) bands. $\Delta$ is introduced as the experimental measure of the degree of dispersion deviation from the bare band. (b) Energy dependence of $\Delta$. Red circles indicate $\Delta$ extracted from $-d^3I/dV^3$, which is from the same dataset used in **Fig. 3a** and **c**. In addition, the peak extraction from curve fitting using multiple Lorentzian [Eq. (1)] on $dI/dV$ is overlaid as gray symbols. See **Appendix A-2 and A-3** for further details on peak extraction processes. (c) Relation between $|\Delta|$ and $|E - E_F|$ (pink symbol) with Lerp $\Delta$ (red curve), which is linearly interpolated $\Delta$. In (c), for simplicity, $\Delta$ from $-d^3I/dV^3$ (and its Lerp $\Delta$) is selectively displayed. The dotted gray line is the linear fitting curve focusing near $E_F$. (d) Energy evolution of LL half width at half maximum (HWHM), which is obtained from the curve fitting using multiple Lorentzian [Eq. (1)]. $E_{kink}$ remarks on the range of the kink around $|E - E_F| \sim 10$ meV. The blue curve is the Kramers-Kronig (KK) transformed Lerp $\Delta$. The error bars in **(b)** and **(d)** correspond to fitting errors obtained using Eq. (1). (e) The Fermi surface of LaTe$_3$ with CDW-driven band folding (green). For simplicity, the ungapped Fermi surface portion near the BZ boundary is selectively shown. (f) Effective Fermi pockets of LaTe$_3$, inspired by LL analysis. An arrow indicates a CDW propagation vector, seen in a non-equilibrium state in LaTe$_3$ [20] (see the main body for details).

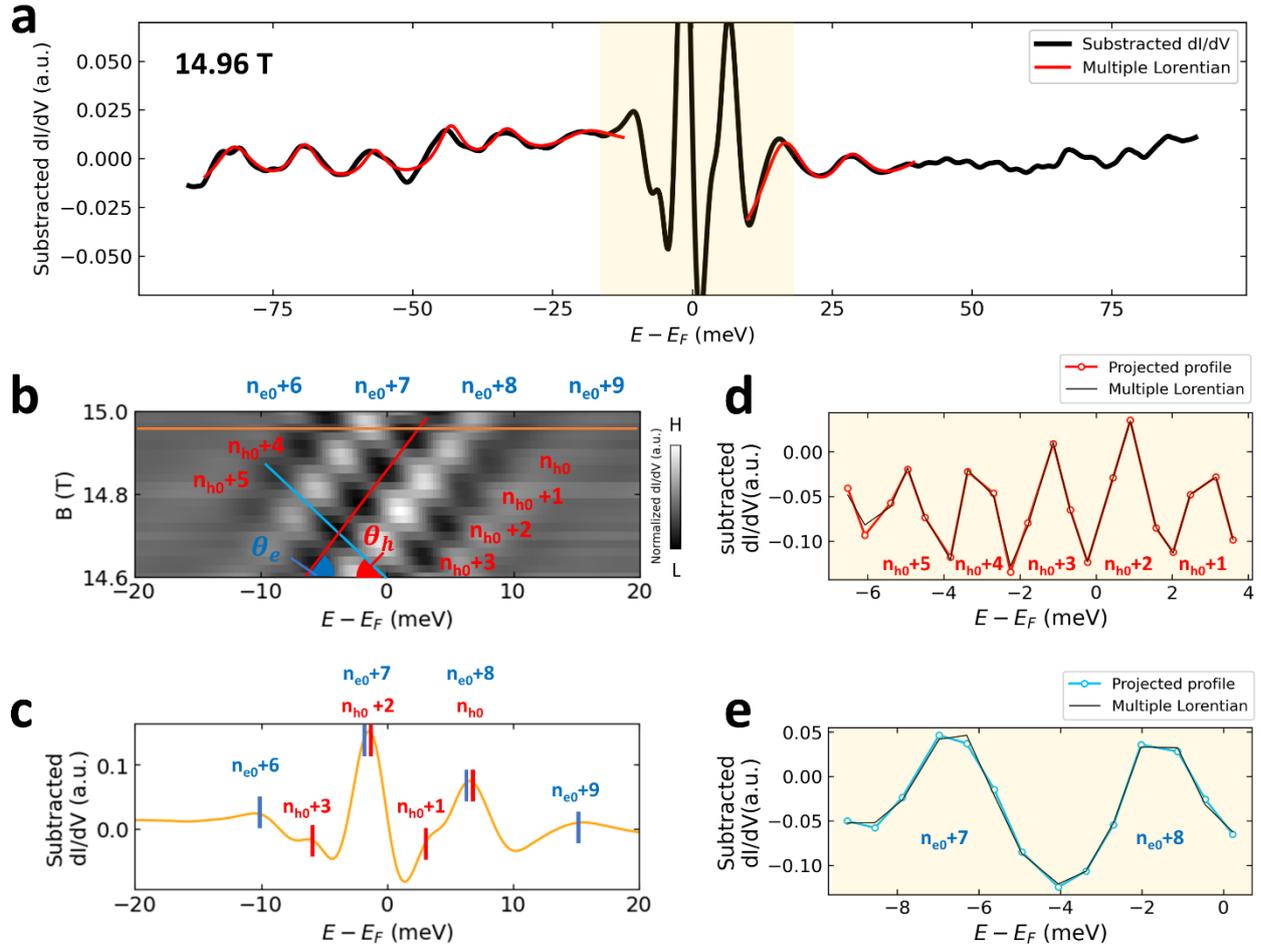

**Figure 5**

(a) The normalized d$I$/d$V$ under 14.96 T (black) and multiple Lorentzian-based fitted curves (red). The shading area is added to represent two distinct energy regions for using different approaches to extract LLs' line widths (see the main body for details). (b) The image plot of d$I$/d$V$ near $E_F$ with a changing magnetic field. Displayed LLs derived from the electron (hole) band are represented with the offset index $n_{e0}$ ($n_{h0}$). (c) The normalized d$I$/d$V$ focusing near $E_F$ at 14.96 T [orange line in (b)]. The blue (red) vertical bars in (c) indicate LLs derived from the electron (hole) band. (d)-(e) The two line profiles along the diagonal lines in (b), which are projected onto the energy axis. See **Appendix A-3** for further details.

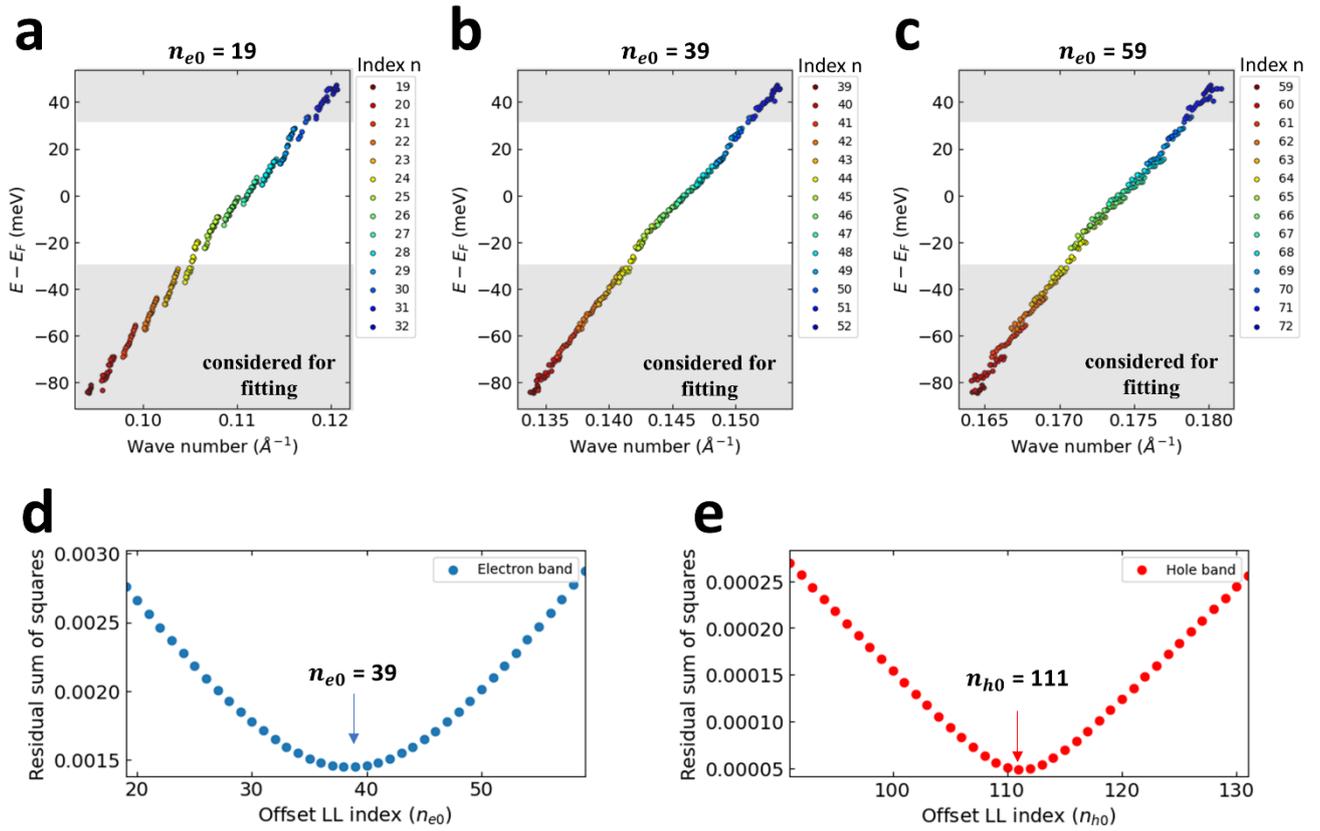

**Figure 6**

(a)-(c) The energy-momentum dispersions with the offset LL index (a) $n_{e0}$ = 19, (b) 39, (c) 59. (d)-(e) The evolution of residual sum of squares with the different offset LL indices (d) $n_{e0}$ and (e) $n_{e0}$. Note that for fixed $n_{e0}$ (and $n_{h0}$), the variation of the residual sum of squares $s$ can be calculated by changing parameters ($c_0$, $c_2$, $c_4$) in Eq. (3). In (d) and (e), thus explored minimum $s$ is plotted as a function of $n_{e0}$ (and $n_{h0}$). See main body and **Appendix B** for further details.

|  | $k_F$ (Å$^{-1}$) | $A_{FS}$ (%BZ) | $E_b$ (eV) | $v_F$ (10$^6$ m/s) | $m^*/m_e$ | $\lambda$ |
|---|---|---|---|---|---|---|
| β (e$^-$) | 0.146(5) | 3.39(1) | -0.85 | 0.80(6) | 0.08(1) | 0.19(1) |
| γ (h$^+$) | 0.227(1) | 8.00(1) | - | - | - | - |
| literature β (e$^-$) | - | 2.44(1) [40] | -0.4~-0.8 [22] | - | 0.175(1) [40] | - |
| literature γ (h$^+$) | - | 7.65(5) [40] | - | 1-1.2 [22] | - | - |

Table 1

Physical properties of LaTe$_3$ obtained in this study. This table includes averaged Fermi wavelength relative to the center of Fermi pocket ($k_F$), area of Fermi surface to Brillouin Zone ($A_{FS}$), energy of band edge [$E_b$, $c_0$ in Eq. (3)], Fermi velocity ($v_F$), ratio of effective mass ($m^*/m_e$), and electron-phonon coupling constant ($\lambda$). Several parameters for γ-band are not shown due to the unresolved bare band in our study. It should be noted that our effective mass (band effective mass) and the previous report (cyclotron effective mass) are not exactly the same. Therefore, the comparison between our results and previous results in $m^*$ is not straightforward. See **Appendix B-3** for further details. The errors in this table originate from the uncertainties in the fitting parameters of Eq. (3) and are calculated based on Eq. (5), (6), and (7).

pockets, as in this study, LL spectroscopy proves to be the most effective method for conveying the key findings.

[50] K. Momma and F. Izumi, VESTA 3 for Three-Dimensional Visualization of Crystal, Volumetric and Morphology Data, Journal of Applied Crystallography **44**, 1272 (2011).

[51] A. Savitzky and M. J. E. Golay, Smoothing and Differentiation of Data by Simplified Least Squares Procedures, Analytical Chemistry **36**, 1627 (1964).

[52] N. W. Ashcroft and N. D. Mermin, 1976, Solid State Physics (Holt, Rinehardt and Winston, New York).25